\documentclass[doublecol]{epl2}
\usepackage{epsfig}
\usepackage{color}
\usepackage{amsmath}
\usepackage{amssymb} 
\title{Interaction of two-dimensional plasma crystals with upstream charged particles}
\shorttitle{}

\author{C.-R. Du, V. Nosenko, S. Zhdanov, H. M. Thomas, and G. E. Morfill}
\shortauthor{}

\institute{Max Planck Institute for Extraterrestrial Physics, Giessenbachstra{\ss}e, 85748 Garching, Germany}


\abstract{
Two-dimensional plasma crystals are characterized by a strong up-and-down asymmetry not only due to gravity 
but also due to the presence of plasma flow at the location of particles. 
We study for the first time the interaction of a single-layer plasma crystal with charged extra particles located
above it (upstream of the flow of ions). 
Upstream extra particles tend to move between the rows of particles in the crystal,
accelerate to supersonic speeds, and excite attraction-dominated Mach cones and wakes in the crystal.}

\begin{document}

\maketitle

\section{Introduction}
\label{sec:intro}

A complex plasma is a weakly ionized gas containing ions, electrons, as well as small solid particles \cite{Fortov:2004,Morfill:2009}.
The particles are usually negatively charged due to the high thermal speed of electrons.
The charge of particles ranges from a few hundreds to several thousands of elementary charges 
depending on the particle size and discharge conditions.
In ground-based experiments the particles can levitate in the pre-electrode area against gravity by a strong electric field. 
Under certain experimental conditions they can be confined in a single layer and self-organize in a triangular lattice with hexagonal symmetry.
Such system is known as 2D plasma crystal.
The in-plane interaction between particles in this system is well described by screened Coulomb (Yukawa) potential
with the screening length defined mainly by electrons \cite{Homann:1997,Rosenberg:1997,Ivlev:2005}.

Since the discovery of plasma crystals \cite{Chu:1994,Thomas:1994,Hayashi:1994}, 
various experiments have been performed with 2D plasma crystals including melting \cite{Nosenko:2008,Nosenko:2009}, 
recrystallization \cite{Knapek:2007}, defect transport \cite{Nosenko:2007}, etc. 
In the course of such experiments, 
one often observes \textit{extra} particles apart from the main 2D lattice layer after injecting particles into plasma.
Those particles, which can be agglomerates or contaminations, 
sometimes move at a high speed, disturb the lattice, and can easily spoil the desired experiments by creating 
wave patterns within the lattice layer.
When the extra particle speed is higher than the sound speed of the lattice,
the disturbance forms a Mach cone.
The Mach cones and wakes associated with extra particles moving beneath the lattice layer were well studied in the past decade \cite{Samsonov:1999,Samsonov:2000,Dubin:2000,Schweigert:2002, Havnes:2002}.
This phenomenon can be used for diagnostic purpose.
In fact, by measuring the angle of a Mach cone and the speed of the source of disturbance, 
the sound speed of the crystalline lattice can be directly estimated.
This method was used for 2D as well as 3D plasma crystals \cite{Havnes:2002,Jiang:2009,Schwabe:2011}.
Besides measuring the sound speed of a plasma crystal,
one can also use the extra particles to heat the crystalline lattice \cite{Nunomura:2005}.

However, in many cases one needs an undisturbed 2D plasma crystal to perform some delicate experiments, 
e.g., to investigate the dynamics of a perfect crystal.
It is relatively easy to get rid of the extra particles beneath the lattice layer (downstream of the flow of ions).
In practice one can drop those extra particles on the bottom electrode by reducing the discharge power at higher pressure.
We call this process ``purification''.

In the experiments performed in our laboratory, Mach cones and related wakes were sometimes observed in 2D plasma crystals even after purification. 
Such wave patterns happened to be induced by extra particles located \emph{above} the lattice layer (upstream of the flow of ions), 
and showed many different features. 

In this Letter we report for the first time on the observation of channeling and leapfrog motion of upstream extra particles, 
accompanied by the excitation of attraction-dominated wakes in the lattice. 

\begin{figure}
\includegraphics[width=0.45\textwidth, bb=0 70 800 550]{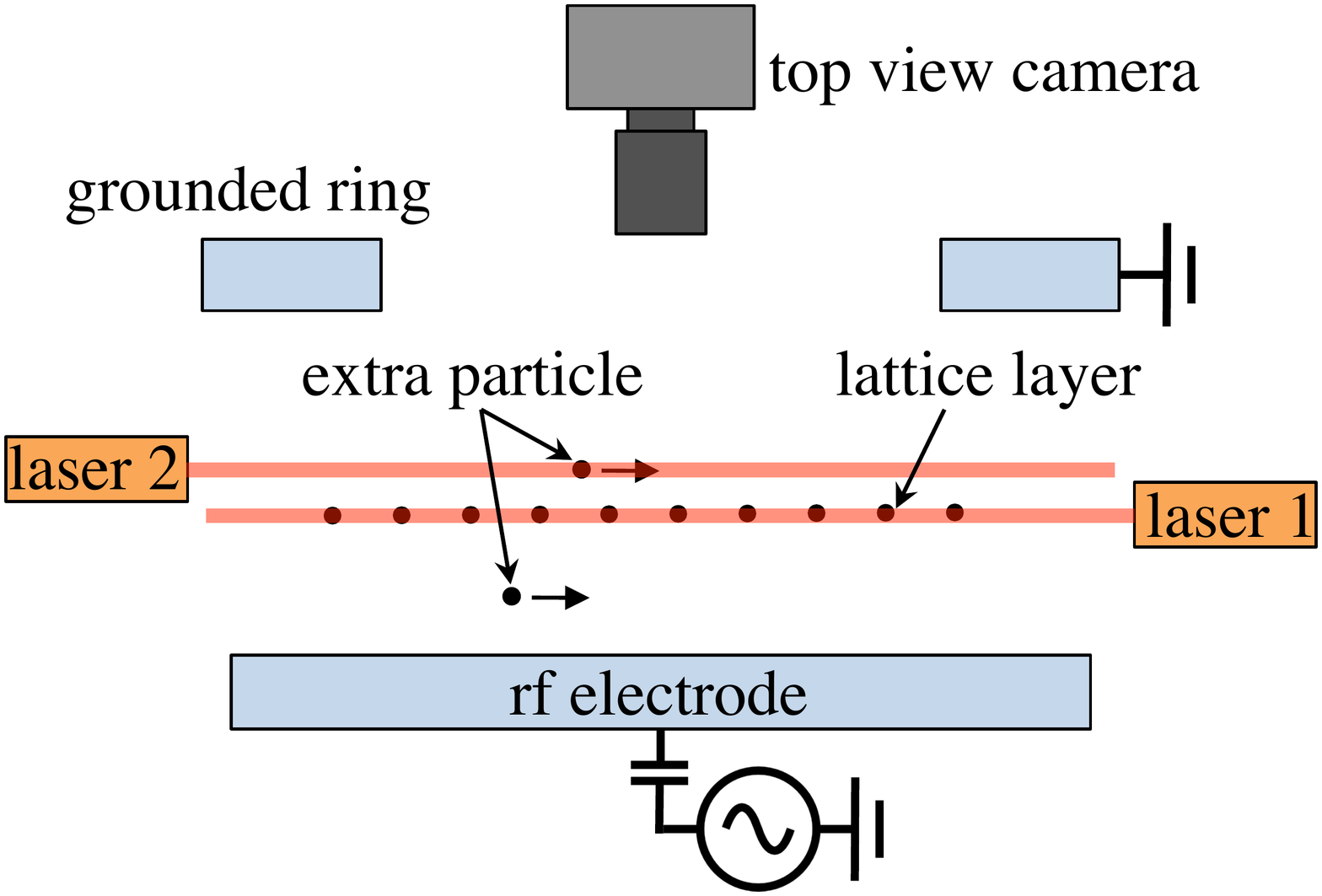}
\caption{Sketch of experimental setup with a modified GEC chamber.
The bottom electrode is powered by rf generator at $13.56$~MHz,
the upper grounded ring and the chamber walls (not shown here) serve as the counter-electrode.
Laser~1 shines a horizontal laser sheet to illuminate the lattice layer of 2D plasma crystal while laser~2 illuminates an extra particle outside the lattice layer.
The optical axes of the lasers are actually oriented at $90^\circ$ to each other.
Particle motion in the lattice layer as well as of the extra particle can be simultaneously recorded by a video camera with top view.
\label{fig:sketch}}
\end{figure}

\section{Experimental setup and conditions}
\label{sec:setup}

The experiments were performed in a modified Gaseous Electronics Conference (GEC) rf reference cell \cite{Nosenko:2012}, see Fig.~\ref{fig:sketch}.
Argon plasma was sustained using a capacitively coupled rf discharge at $13.56$~MHz.
The input power was set at $20$~W.

We used monodisperse melamine formaldehyde (MF) and polystyrene (PS) particles to create 2D plasma crystals suspended above the bottom rf electrode.
The MF particles have a diameter of $9.19\pm0.09$~$\mu$m and mass density of $1.51$~g/cm$^{3}$,
while the PS particles have a diameter of $11.36\pm0.12$~$\mu$m and mass density of $1.05$~g/cm$^{3}$.
Gas pressure was maintained at about $0.65$~Pa; 
the corresponding neutral gas damping rate was $\gamma\simeq0.77$~s$^{-1}$ for MF particles and $\gamma\simeq0.91$~s$^{-1}$ for PS particles \cite{Liu:2003}.
Further experimental parameters are listed in Table~\ref{table:parameters}.
The lattice layer was illuminated by a horizontal laser sheet shining through a side window of the chamber.
A high-resolution video camera (Photron FASTCAM~1024~PCI) was mounted above the chamber, 
capturing a top view with a size of $42\times42$~mm$^2$, as sketched in Fig.~\ref{fig:sketch}.
The recording rate was set at $60$~frames per second.

\begin{largetable}
\caption{Experimental parameters$^a$ including the particle material 
and diameter $d$, charge number $Z=-Q/\textit{e}$, interparticle distance$^b$ $\Delta$, screening length $\lambda_D$,
longitudinal sound speed $C_L$, extra particle speed $v_d$, Mach number $M$, and Mach cone type (see text for details).
The particles were suspended in an argon discharge at the pressure of $0.65$~Pa and discharge power of $20$~W.}
\label{table:parameters}
\begin{tabular}[bt]{c|cccccc|cc|cc}
\hline \hline
experiment	& particle	& $d$		& $Z$		& $\Delta$	&$\lambda_D$	& $C_L$		& extra			& $v_d$		& $M$	& Mach cone		\\
			& material	& ($\mu$m)	&			& ($\mu$m) 	& ($\mu$m) 		& (mm/s)	& particle		& (mm/s)	&		& type			\\
\hline  
 $1$		& MF 		& $9.19$ 	& $15400$ 	& $520$		& $390$			& $27$  	& upstream 		& $32$	 	& $1.2$	& I		\\
 $2$		& MF 		& $9.19$ 	& $16300$ 	& $560$		& $600$			& $33$ 		& downstream 	& $42$ 		& $1.3$	& II 		\\
 $3$		& MF 		& $9.19$ 	& $15000$ 	& $610$		& $300$			& $17$ 		& upstream 		& $24$ 		& $1.4$	& I 		\\
 $4$		& MF 		& $9.19$ 	& $16100$ 	& $560$		& $400$			& $26$ 		& downstream 	& $42$ 		& $1.6$	& II 		\\
 $5$		& PS 		& $11.36$ 	& $19000$ 	& $650$		& $560$			& $27$ 		& upstream 		& $29$ 		& $1.1$	& I			\\
\hline \hline
\multicolumn{11}{l}{\footnotesize{$^a$ Error bars: $\pm13$\% for $Z$,  $\pm45$\% for $\lambda_D$, $\pm15$\% for $C_L$, $\pm5$\% for $v_d$, and $\pm15$\% for $M$. $^b$ $\Delta$ is obtained from the first peak of}} \\
\multicolumn{11}{l}{\footnotesize{the pair correlation function. The crystalline lattice is slightly inhomogeneous ($2$\% at the most).}} 
\end{tabular}
\end{largetable}

\section{Mach cones and wakes}
\label{sec:mach}

\begin{figure}
\includegraphics[width=0.45\textwidth, bb=0 0 700 1100]{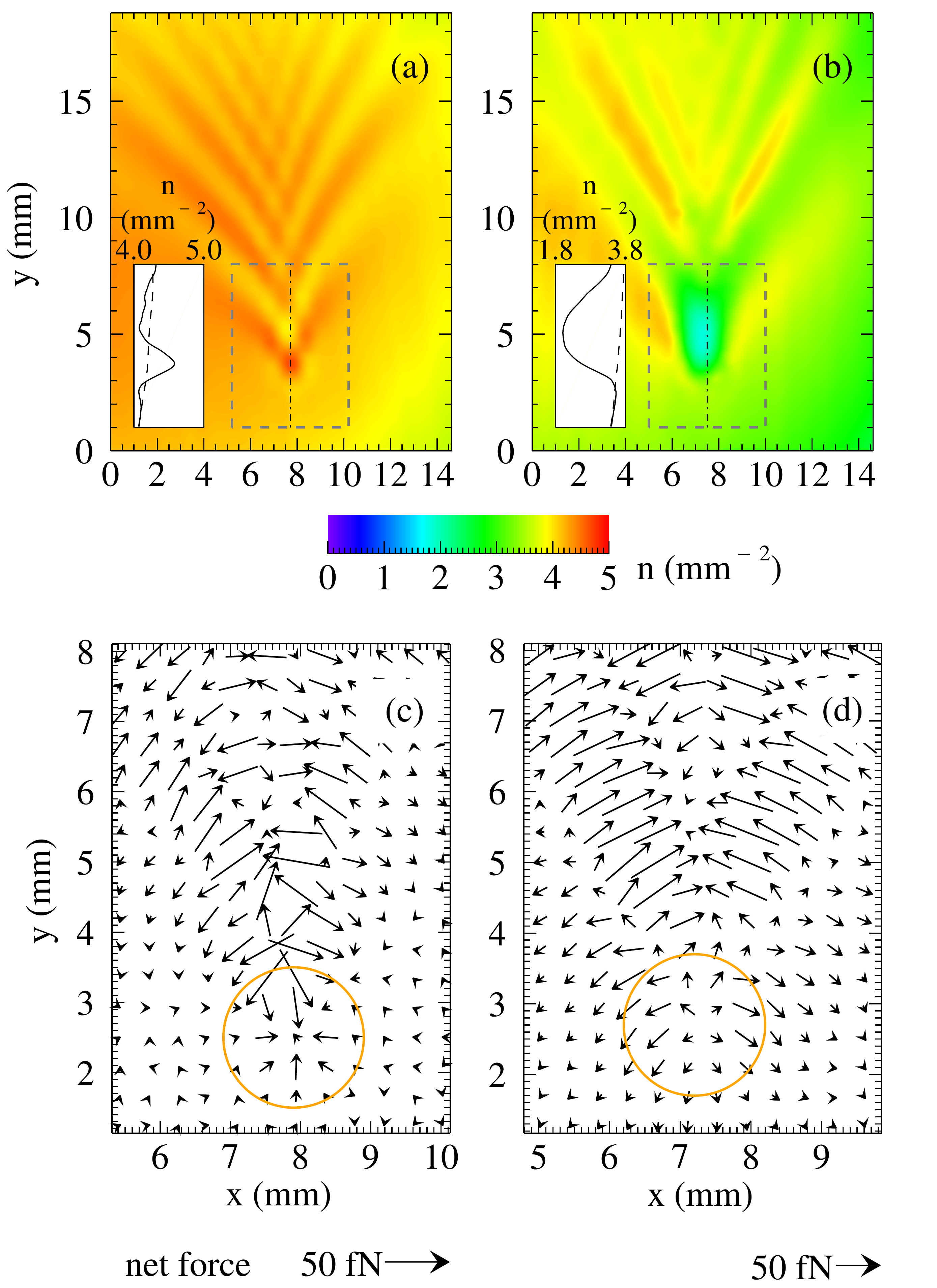}
\caption{Comparison of the particle areal number density $n$ and net force between (a), (c) type I 
(experiment~$1$ in Table~\ref{table:parameters}) and (b), (d) type II (experiment~$2$ in Table~\ref{table:parameters}) Mach cones.
The maps for type I and type II Mach cones are averaged from data for $30$ and $20$ consecutive video frames, 
respectively.
The insets show the density profile (solid lines) along the symmetry axis (marked by the dash-dotted lines).
The dashed lines in the insets represent the density profile for the undisturbed lattice.
The net force $\mathbf{F}_n=m\mathbf{a}+{\gamma}m\mathbf{v}$ acting on the particles in the lattice at the apex of the Mach cone [marked by the dashed rectangle in (a) and (b)] is shown in panels (c) and (d). 
The orange circles highlight the apex of the Mach cones. 
\label{fig:machcone}}
\end{figure}

We performed two separate experiments intended to observe two different types of Mach cones.
For the sake of simplicity, 
in this Letter we name the Mach cones in the lattice observed after its purification the ``type I'' Mach cones.
In contrast, 
the well-known Mach cones excited by extra particles beneath the lattice layer are called the ``type II''.
To observe the type II Mach cones, we did not purify the plasma crystal so that big extra particles remained in the discharge.
To make the results comparable, two cases with the similar Mach numbers\footnote{
The Mach number is defined as $M = v_d/C_L$, 
where $v_d$ is the speed of disturbance (extra particle) and $C_L$ is the longitudinal sound speed of the crystalline lattice.}
but related to two different excitation sources were selected, see Fig.~\ref{fig:machcone}. 
The particle areal number density map and the net force vector field plot corresponding to the type I Mach cone are shown in the left panels in Fig.~\ref{fig:machcone}.
The wings of the Mach cone are clearly recognizable in both plots.
The half-opening angle between the wings (``cone angle'' in what follows) is measured to be $\mu = 55^{\circ}\pm10^{\circ}$. 
By using the well-known Mach cone relation $\sin \mu = M^{-1}$, where $M > 1$, 
one can obtain $M\simeq1.2$. 
The measured Mach number agrees well with the  value estimated using the phonon spectrum method \cite{Nunomura:2002}.
Note that the lateral wakes are also well resolved in the density map; 
these can be used for diagnostic purpose \cite{Nosenko:2003}.

In the density map, Fig.~\ref{fig:machcone}(a), 
there is a density increase at the apex of the type I Mach cone. 
Comparing to the undisturbed value\footnote{
Note that the unperturbed crystalline lattice inhomogeneity is small compared to the density variation caused by the extra particle.} 
the particle density increases from $4$~mm$^{-2}$ to $4.5$~mm$^{-2}$. 
The reason for that is simple: 
The particles in the lattice layer are dragged toward each other, 
resulting in an increase of the local particle density. 
This is clearly seen in the net force vector field shown in Fig.~\ref{fig:machcone}(c).

As to the ``conventional'' type II Mach cone, 
the density at the apex drops down sharply from  $3.7$~mm$^{-2}$ (undisturbed particle density) to $1.9$~mm$^{-2}$, 
producing a hole in the lattice clearly seen in Fig.~\ref{fig:machcone}(b). 
The density drops because the lattice particles at the apex of the Mach cone are repelled by the extra particle, see  Fig.~\ref{fig:machcone}(d). 
The repulsion is caused by the Yukawa repulsive force between the extra particle and the particles in the lattice layer, 
all of which are negatively charged. 
The cone angle\footnote{Measured in the particle speed map (not shown here).} is $\mu= 50^{\circ}\pm6^{\circ}$.

\section{Extra particle diagnostics}
\label{sec:extra}

\begin{figure}
\includegraphics[width=0.48\textwidth, bb=0 0 900 950]{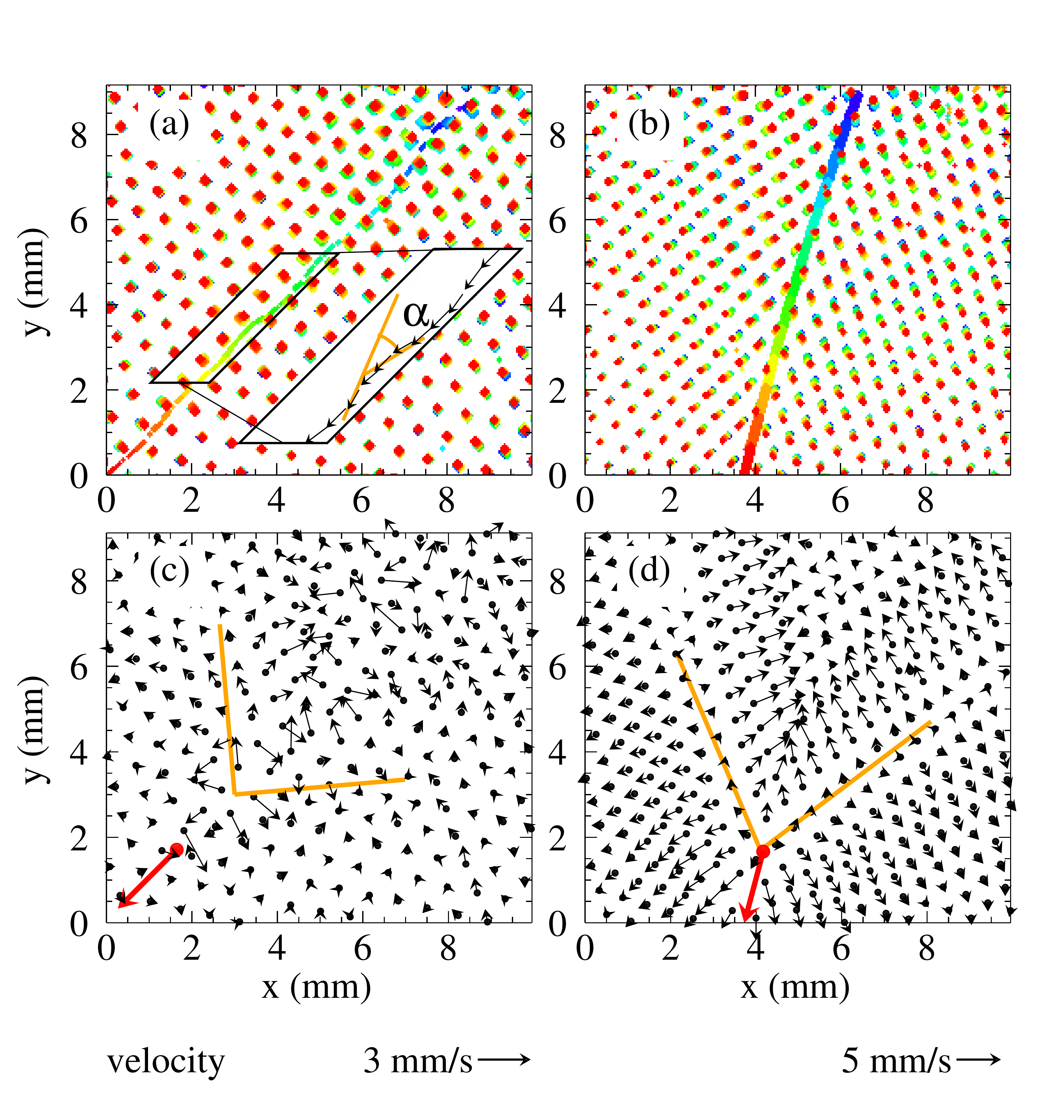}
\caption{Trajectories of
upstream (a) and downstream (b) extra particles (experiments~$3$ and $4$ in Table~\ref{table:parameters}).
The particle positions are presented by superimposing a series of consecutive experimental images for $0.45$~s (a) and $0.22$~s (b)
(color-coded from blue to red).
The trajectory of the upstream particle, showing a clear zig-zag feature, is magnified in the inset in (a).
The scattering angle is $\alpha=25^\circ\pm3^\circ$.
Panels (c) and (d) show snapshots of positions (black dots) and velocities (black scaled arrows) of particles in the lattice layer in a single frame as well as positions (red dots) of upstream and downstream particles and their velocities (red arrows, scaled $1:5$), respectively.
In (c) and (d) the Mach cone is highlighted by orange lines.
The cone angles are $44^\circ\pm12^\circ$ (c) and $38^\circ\pm8^\circ$ (d).
\label{fig:extra}}
\end{figure}

In order to trace the extra particle originating the Mach cone,
we installed a second laser at $90^\circ$ to the first one.
This laser shines a laser sheet, which is parallel to the first laser sheet, as shown in Fig.~\ref{fig:sketch}.  
The height of the laser can be finely adjusted with accuracy of $10$~$\mu$m.
Using two lasers simultaneously and registering the light scattered by the particles with the same camera allowed us to image the lattice layer together with the extra particle.
(The depth of field of the camera lens is larger than the distance between the extra particle and the lattice layer.)
The trajectory of the extra particle is therefore overlapped with the trajectories of particles in the lattice layer, see Figs.~\ref{fig:extra}(a), (b). 
That fits well with our goals.

The extra particles beneath the lattice layer moved at much lower heights. 
They could be observed if we lowered the height of the second laser.
For brevity, we call these particles ``downstream'' because they are located downstream of the ion flow in the (pre)sheath comparing to the particles in the lattice layer.
As shown in Fig.~\ref{fig:extra}(b),
the trajectory of this extra particle beneath the lattice layer is well resolved as well as the displacement of individual particles in the lattice layer.
The extra particle trajectory is rather smooth and apparently not influenced by the local structure of the lattice layer above it.
By measuring the height difference between two laser sheets, 
we estimate the vertical distance of extra particle from the lattice layer to be  $\Delta h \simeq 800$~$\mu$m.
The Mach cone can already be seen in Fig.~\ref{fig:extra}(b).
However, the cone structure is better resolved in the velocity vector field, Fig.~\ref{fig:extra}(d).
It is clear that the extra particle repels the particles in the lattice layer right above it due to the Yukawa repulsion.
The apex of the Mach cone follows this extra particle closely.

It is more difficult to find the source of type I Mach cone.
We scan the second laser vertically with small height steps to catch the trace of any possible source of disturbance.
It turns out that the source of type I Mach cone is an extra particle moving above the lattice layer.
For the reason explained above, we call these particles ``upstream''.
The relative height of these particles above the lattice layer is estimated to be  $\Delta h \simeq 200$ ~$\mu$m.
There are several reasons why it is hard to visualize or in fact even notice the existence of such particles:
(i) This type of extra particles appears in a plasma crystal much less often than downstream particles,
(ii) these particles are generally smaller in size, 
resulting in a very dim trajectory recorded by the video camera,
(iii) the influence of such particle on the lattice layer is relatively small. 
As we already see in the density maps in Fig.~\ref{fig:machcone}, 
the density variation caused by an upstream particle is much smaller then that caused by downstream particle.
Thus it is easy to overlook such effect without careful analysis.

As a rule, the upstream particles appear to move slower than downstream particles.
Their average velocity depends on the gas pressure, discharge power, particle size, etc.
Such dependence is not the focus of this Letter and will be reported elsewhere.

\section{Extra particle -- lattice interaction at the cone apex}
\label{sec:inter}

The computed velocity of particles in the lattice layer with Mach cones excited by upstream and downstream particles are shown in Fig.~\ref{fig:extra}(c) (type I) and Fig.~\ref{fig:extra}(d) (type II), respectively. 
It is immediately clear that they are strikingly different.  
First, in the vicinity of the apex of type I cone, 
the lattice layer particles move toward the upstream particle [Fig.~\ref{fig:extra}(c)], 
whereas they are visibly repelled by the downstream particle [Fig.~\ref{fig:extra}(d)]. 
Second, the apex of the Mach cone in Fig.~\ref{fig:extra}(c) has a elongated shape along the direction of the motion of the extra particle. 
The conjunction of the wings of the cone is therefore located slightly behind the moving upstream particle at a distance $\sim (2-3)\Delta$. 
In contrast, the apex of the type II cone follows the extra particle position closely, as shown in Fig.~\ref{fig:extra}(d).

Importantly, the force field in Figs.~\ref{fig:machcone}(c),(d) and the velocity field in Figs.~\ref{fig:extra}(c),(d) are similarly ``polarized'', 
i.e., both fields are simultaneously directed toward or away from the instantaneous location of the extra particle. 
Based on what we observed, the extra particle -- lattice layer interaction is dominantly \emph{attractive} for type I cone (upstream particle) or dominantly \emph{repulsive} for type II cone (downstream particle).

We attribute the particle-lattice layer attraction to the \emph{ion wake} formed underneath the upstream extra particle\footnote{
As is the case for any particle levitating in a discharge (pre)sheath.}. 
Positive ions that concentrate locally below the extra particle exert an attractive force on the negatively charged particles of the lattice layer. 
This phenomenon is well known as the ion-wake effect causing, e.g. the particles to pair \cite{Lampe:2000,Melzer:2000,Morfill:2009,Io:2010,Kroll:2010}.
The particle-lattice repulsion resulting from the Yukawa repulsive force at the apex of the type II cone is a well-established effect \cite{Samsonov:1999,Samsonov:2000,Dubin:2000,Schweigert:2002, Havnes:2002}.

Finally, the situation appears to be more complicated, and the particle -- lattice interaction cannot be reduced only to the dominantly-attractive or dominantly-repulsive interactions. 
The examples discussed below can be properly addressed only suggesting the presence of competing repulsive-attractive interactions 
(which are also considered to be important for colloids \cite{Reichhardt:2004}.)

\section{Channeling effect}
\label{sec:chan}

There is one more surprising observation regarding upstream particles: 
The long-term behavior of these particles demonstrates anomalous transport properties and elements of ``strange kinetics'' \cite{Shlesinger:1993}. 
For instance, 
the particle is apparently able to channel\footnote{
Channeling is a process that constrains the path of a charged particle in a crystalline solid \cite{Feldman:1982}. 
} 
between two aligned rows of particles in the crystalline layer (``wall particles'' below),
as seen in Fig.~\ref{fig:extra}(a).

\subsection{Confinement force}
\label{subsec:confine}

The upstream particle is well confined in the channel. 
Comparing to the smooth trajectory of the downstream particle, Fig.~\ref{fig:extra}(b), 
the trajectory in Fig.~\ref{fig:extra}(a) has a zig-zag shape [see inset where a part of the trajectory in question is magnified]. 
In other words, the extra particle, interacting with the crystal particles comprising the channel, 
evidently bounces and hence the effect of the wall particles on the extra particle is dominated by\emph{ repulsion}.
By measuring the scattering angle $\alpha$, 
one can estimate the confinement force $F_c$ by using the relation $F_c\Delta t = m<v_d>(1 - \cos \alpha)$, 
implying momentum conservation, 
where $\Delta t \simeq0.08$~s is the time of scattering, 
$m$ and $<v_d>$ are the mass and mean (longitudinal) velocity of the extra particle, respectively. 
In the experiment presented here, the angle is measured to be $\alpha =25^{\circ}\pm3^{\circ}$, 
$v_d= 29\pm2$~mm/s, and the confinement force is estimated as $F_c \simeq 21$~fN. 
This value is about twice the neutral gas friction force $F_n \simeq 13$ fN.

\subsection{Non-reciprocal interaction}
\label{subsec:nonrecip}

The extra particle, passing through the channel, in turn exerts the force on the wall particles. 
This force deforms the lattice cells when the extra particle moves through. 
Studying this deformation, a certain conclusion can be made about the wall particle -- extra particle interaction. 
From  Fig.~\ref{fig:extra}(a) it follows immediately that  this interaction is \emph{non-reciprocal} 
because the wall particles behave as if they are attracted to rather than repelled by the similarly charged extra-particle. 
This kind of non-reciprocity is indeed easy to address taking into account the ion wake (ion focus \cite{Morfill:2009}) formed beneath the upstream particle.

\begin{figure}
\includegraphics[width=0.45\textwidth, bb=0 0 650 900]{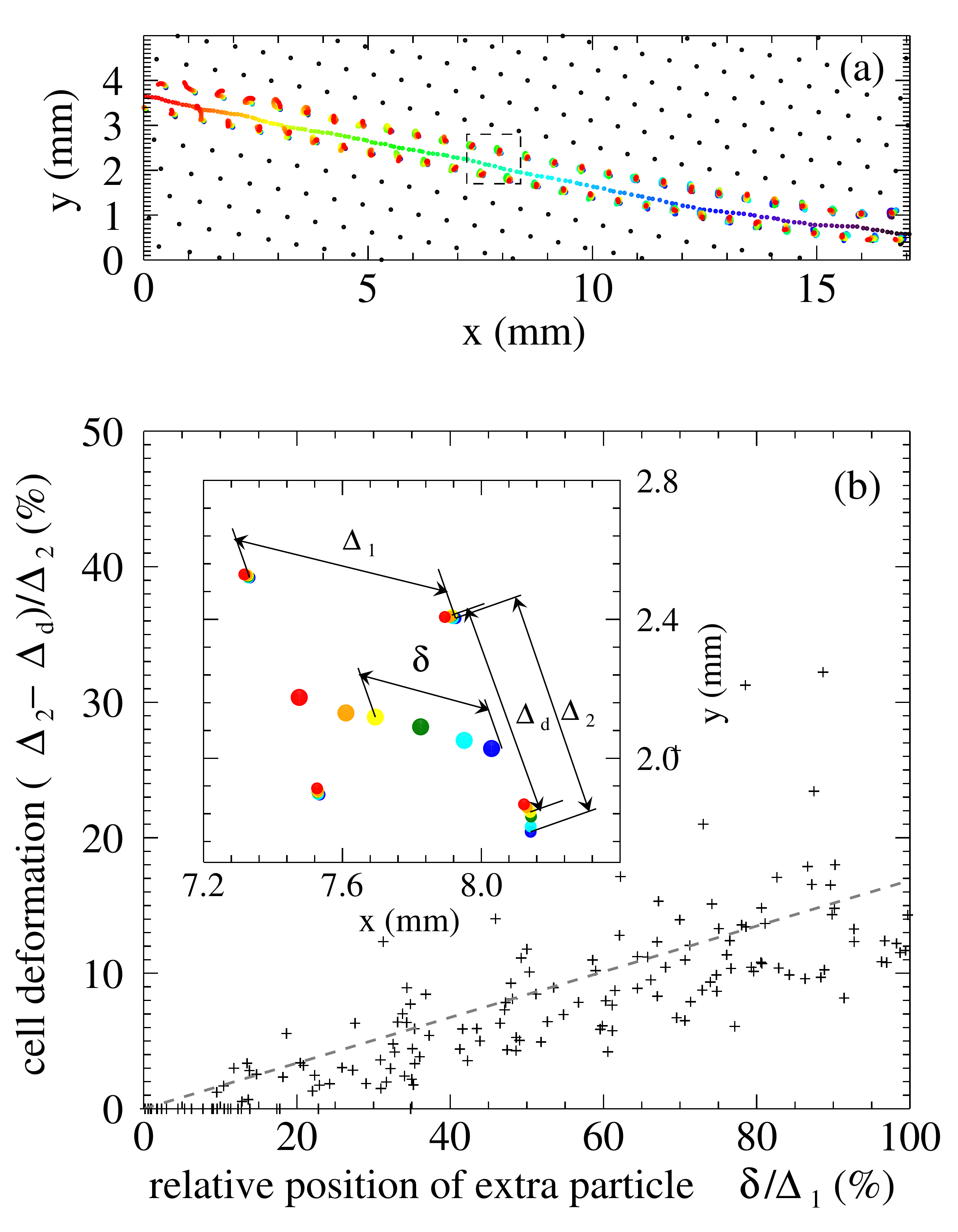}
\caption{Channeling of an upstream particle (a) and related cell deformation (b) (experiment~$5$ in Table \ref{table:parameters}).
Panel (a): black dots represent the lattice particles, colored dots represent positions of the extra particle and the ``wall'' particles color-coded from blue to red for $0.66$~s.
Panel (b), inset: A magnified single cell [marked by a dashed-line rectangle in (a)] with color-coded particle positions (from blue to red with a time step of $4$~ms).
The dashed line is the least squares linear fit.
\label{fig:channeling}}
\end{figure}

\subsection{Lattice cell deformation}
\label{subsec:deform}

In order to investigate the deformation of the lattice cell caused by the passage of extra particle in more detail, 
we performed another experiment with higher recording rate of 250 frames per second (experiment~$5$ in Table~\ref{table:parameters}). 
Accordingly, we used particles with larger diameter 
so that they can be well illuminated with the same illumination laser power without a significant change of the recording quality. 
An (unexpected) advantage of this case is that the extra particle above the lattice layer can even be traced using the same laser. 
This implies that the height difference is roughly of the laser sheet width or even smaller, $\Delta h \leq 100$ $\mu$m. 
This height difference is about one-fifth of the characteristic length $L_E = E/E_h'$ at the levitation height in the (pre)sheath, 
where $E$ is the local electric field \footnote{
Note that both PS and MF particles levitated at approximately the same height.}. 
The expected difference in particle mass $\Delta m/m\approx 1.5 \Delta h/L_E$ is less than $30$--$35$\%.

\begin{figure}
\includegraphics[width=0.42\textwidth, bb=0 20 500 700]{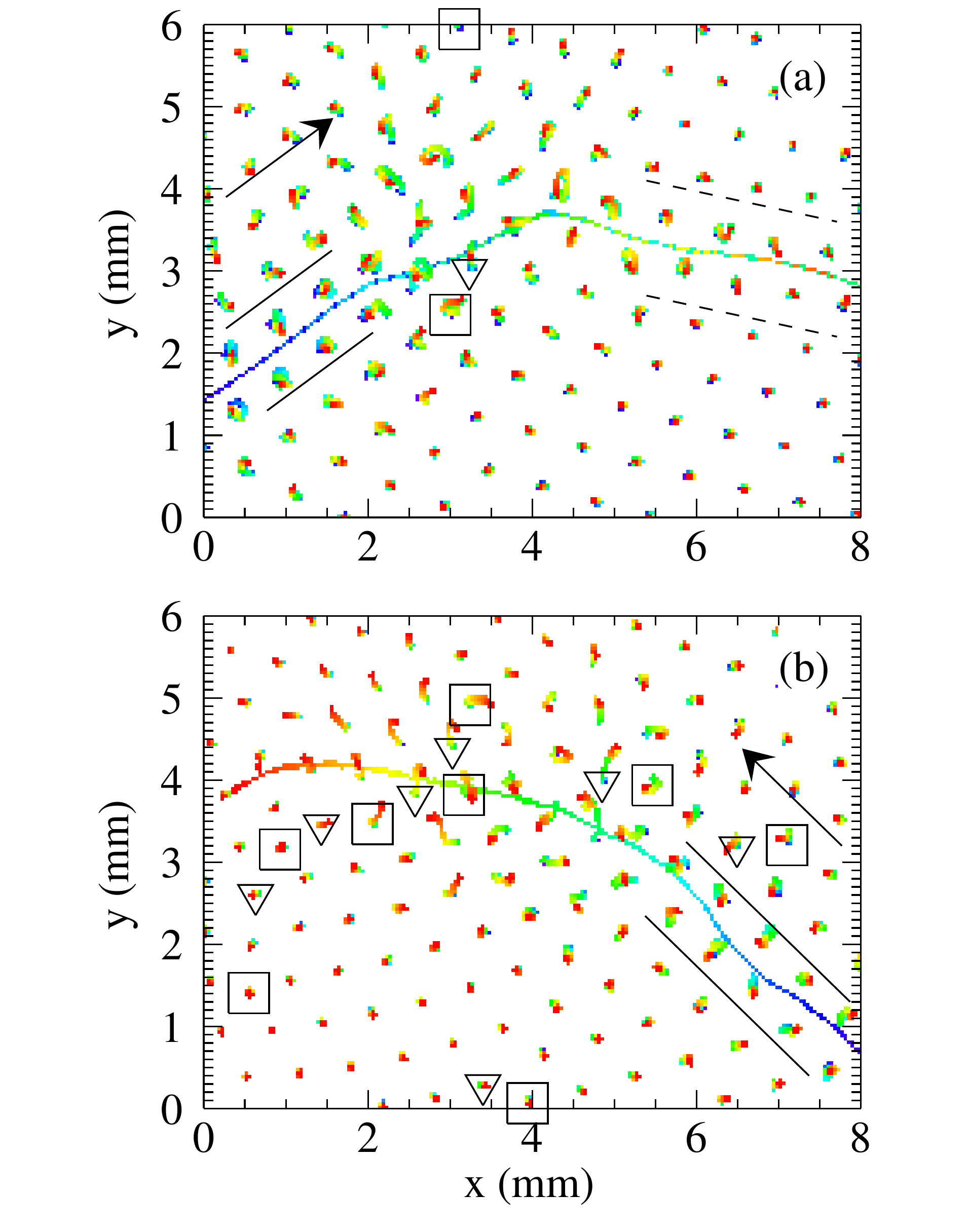}
\caption{Interaction of upstream particles with defects in the crystalline lattice. 
The particle positions obtained for the time interval of 0.4 s are shown as the solid dots color-coded from blue to red. 
Initially in both cases the extra particles move in the channel (highlighted by two parallel solid lines). 
The initial velocities are approximately the same $v_d = 26\pm1$ mm/s. 
(a) -- single point defect scattering; 
the scattered particle is captured in a new channel (highlighted by two parallel dashed lines) 
where it moves with the velocity $v_d = 27\pm1$ mm/s. 
(b) -- interaction with a large-angle grain boundary;
the extra particle, being initially scattered by a single defect, 
left the channel and penetrated the chain of defects, 
jumping from one position to another with the mean velocity $v_d  = 19\pm3$ mm/s. 
The $7$-fold cells are marked by triangles, and $5$-fold cells are marked by squares.
\label{fig:defect}}
\end{figure}

The extra particle track and the accompanying cell deformation are shown in Fig.~\ref{fig:channeling}. 
The extra particle moves along the channel, 
which even bends slightly at $x > 14$~mm as shown in Fig.~\ref{fig:channeling}(a). 
The deviation from the major track is small and a zig-zag feature is barely visible in this figure, 
indicating stronger confinement as compared to Fig.~\ref{fig:extra}(a). 
Let us focus on the dynamics of a single cell, see the inset in Fig.~\ref{fig:channeling}(b). 
When the extra particle enters (through the right boundary in the case considered) the cell deforms as indicated by the color-coding in the inset. 
In order to quantify this deformation, we introduce the relative cell deformation $(\Delta_2-\Delta_d)/\Delta_2$, 
where $\Delta_{2(d)}$ is the undisturbed (disturbed) length of the cell side transversal to the extra particle track; 
the relative position of the extra particle inside the cell $\delta/\Delta_1$, 
where $\Delta_{1}$ is the length of the side of the undisturbed cell  along the track, 
and $\delta$ is the displacement of the extra particle measured from the right boundary of the cell, as shown in Fig.~\ref{fig:channeling}. 
The deformation depends linearly on the relative position of extra particle: 
$\delta/\Delta_1=k(\Delta_2-\Delta_d)/\Delta_2$, where $k = 0.14 \pm 0.01$. 
Note also that in fact the deformation does not only depend on the relative position of the extra particle, 
but also on its velocity: 
At a fixed  $\delta/\Delta_1$, the slowing-down of the particle leads to a higher deformation of the lattice up to $30$\%,
as seen in Fig.~\ref{fig:channeling}(b).

\subsection{Interaction with lattice defects}
\label{subsec:defect}

The character of the upstream extra particle motion depends on the local structure of the lattice layer. 
Sometimes the extra particles move in irregular trajectories as we observed in a number of experiments
(two examples are shown in Fig.~\ref{fig:defect}). 
This can be explained by the channel distortion. 
For instance, when the channel is blocked by a structural defect the extra particle collides with the defect and scatters leaving the channel. 
Depending on the relative kinetic energy the collision might be elastic or inelastic. 
After the collision the particle is often again involved in the channeling process, see Fig.~\ref{fig:defect}(a), 
being again captured by and accelerated along a channel with a new orientation. 
Sometimes it is a challenge for an extra particle to find a new suitable channel, 
and it remains quasi-free ``leapfrog'' jumping for a longer time from one position to another one but at an essentially lower velocity, 
see Fig.~\ref{fig:defect}(b).
This requires further careful analysis.

To conclude, 
the upstream extra particles happened to be extremely useful tool to effectively test the anomalous kinetics effects and the particle-lattice interaction.
Those particles appeared to be weakly, quasi-elastically interacting with a strongly coupled 2D complex plasma.
That allowed us to explore the anomalous long-term channeling, leapfrog motion,
attraction-dominated wakes in the lattice, all at the ``atomistic'' level.

\acknowledgments
We thank Alexei Ivlev for valuable discussions.
The research leading to these results has received funding from the European
Research Council under the European Union's Seventh Framework Programme
(FP7/2007-2013) / ERC Grant agreement 267499.


\begin{thebibliography}{10}
\expandafter\ifx\csname url\endcsname\relax\def\url#1{\texttt{#1}}\fi

\bibitem{Fortov:2004}
\Name{Fortov V.~E., Khrapak A.~G., Khrapak S.~A., Molotkov V.~I. \and Petrov
  O.~F.} \REVIEW{Physics-Uspekhi }{47}{2004}{447}.

\bibitem{Morfill:2009}
\Name{Morfill G.~E. \and Ivlev A.~V.} \REVIEW{Rev. Mod. Phys.
  }{81}{2009}{1353}.

\bibitem{Homann:1997}
\Name{Homann A., Melzer A., Peters S. \and Piel A.} \REVIEW{Phys. Rev. E
  }{56}{1997}{7138}.

\bibitem{Rosenberg:1997}
\Name{Rosenberg M. \and Kalman G.} \REVIEW{Phys. Rev. E }{56}{1997}{7166}.

\bibitem{Ivlev:2005}
\Name{Ivlev A.~V., Zhdanov S.~K., Khrapak S.~A. \and Morfill G.~E.}
  \REVIEW{Phys. Rev. E }{71}{2005}{016405}.

\bibitem{Chu:1994}
\Name{Chu J.~H. \and I L.} \REVIEW{Phys. Rev. Lett. }{72}{1994}{4009}.

\bibitem{Thomas:1994}
\Name{Thomas H., Morfill G.~E., Demmel V., Goree J., Feuerbacher B. \and
  M\"ohlmann D.} \REVIEW{Phys. Rev. Lett. }{73}{1994}{652}.

\bibitem{Hayashi:1994}
\Name{Hayashi Y. \and Tachibana K.} \REVIEW{Japanese Journal of Applied Physics
  }{33}{1994}{L804}.

\bibitem{Nosenko:2008}
\Name{Nosenko V., Zhdanov S., Ivlev A.~V., Morfill G., Goree J. \and Piel A.}
  \REVIEW{Phys. Rev. Lett. }{100}{2008}{025003}.

\bibitem{Nosenko:2009}
\Name{Nosenko V., Zhdanov S.~K., Ivlev A.~V., Knapek C.~A. \and Morfill G.~E.}
  \REVIEW{Phys. Rev. Lett. }{103}{2009}{015001}.

\bibitem{Knapek:2007}
\Name{Knapek C.~A., Samsonov D., Zhdanov S., Konopka U. \and Morfill G.~E.}
  \REVIEW{Phys. Rev. Lett. }{98}{2007}{015004}.

\bibitem{Nosenko:2007}
\Name{Nosenko V., Zhdanov S. \and Morfill G.} \REVIEW{Phys. Rev. Lett.
  }{99}{2007}{025002}.

\bibitem{Samsonov:1999}
\Name{Samsonov D., Goree J., Ma Z.~W., Bhattacharjee A., Thomas H.~M. \and
  Morfill G.~E.} \REVIEW{Phys. Rev. Lett. }{83}{1999}{3649}.

\bibitem{Samsonov:2000}
\Name{Samsonov D., Goree J., Thomas H.~M. \and Morfill G.~E.} \REVIEW{Phys.
  Rev. E }{61}{2000}{5557}.

\bibitem{Dubin:2000}
\Name{Dubin D. H.~E.} \REVIEW{Physics of Plasmas }{7}{2000}{3895}.

\bibitem{Schweigert:2002}
\Name{Schweigert V.~A., Schweigert I.~V., Nosenko V. \and Goree J.}
  \REVIEW{Physics of Plasmas }{9}{2002}{4465}.

\bibitem{Havnes:2002}
\Name{Havnes O., Hartquist T.~W., Brattli A., Kroesen G. M.~W. \and Morfill G.}
  \REVIEW{Phys. Rev. E }{65}{2002}{045403}.

\bibitem{Jiang:2009}
\Name{Jiang K., Nosenko V., Li Y.~F., Schwabe M., Konopka U., Ivlev A.~V.,
  Fortov V.~E., Molotkov V.~I., Lipaev A.~M., Petrov O.~F., Turin M.~V., Thomas
  H.~M. \and Morfill G.~E.} \REVIEW{EPL (Europhysics Letters)
  }{85}{2009}{45002}.

\bibitem{Schwabe:2011}
\Name{Schwabe M., Jiang K., Zhdanov S., Hagl T., Huber P., Ivlev A.~V., Lipaev
  A.~M., Molotkov V.~I., Naumkin V.~N., S\"utterlin K.~R., Thomas H.~M., Fortov
  V.~E., Morfill G.~E., Skvortsov A. \and Volkov S.} \REVIEW{EPL (Europhysics
  Letters) }{96}{2011}{55001}.

\bibitem{Nunomura:2005}
\Name{Nunomura S., Zhdanov S., Samsonov D. \and Morfill G.} \REVIEW{Phys. Rev.
  Lett. }{94}{2005}{045001}.

\bibitem{Nosenko:2012}
\Name{Nosenko V., Ivlev A.~V. \and Morfill G.~E.} \REVIEW{Phys. Rev. Lett.
  }{108}{2012}{135005}.

\bibitem{Liu:2003}
\Name{Liu B., Goree J., Nosenko V. \and Boufendi L.} \REVIEW{Physics of Plasmas
  }{10}{2003}{9}.

\bibitem{Nunomura:2002}
\Name{Nunomura S., Goree J., Hu S., Wang X. \and Bhattacharjee A.}
  \REVIEW{Phys. Rev. E }{65}{2002}{066402}.

\bibitem{Nosenko:2003}
\Name{Nosenko V., Goree J., Ma Z.~W., Dubin D. H.~E. \and Piel A.}
  \REVIEW{Phys. Rev. E }{68}{2003}{056409}.

\bibitem{Lampe:2000}
\Name{Lampe M., Joyce G., Ganguli G. \and Gavrishchaka V.} \REVIEW{Physics of
  Plasmas }{7}{2000}{3851}.

\bibitem{Melzer:2000}
\Name{Melzer A., Schweigert V.~A. \and Piel A.} \REVIEW{Physica Scripta
  }{61}{2000}{494}.

\bibitem{Io:2010}
\Name{Io C.-W., Chan C.-L. \and I L.} \REVIEW{Physics of Plasmas
  }{17}{2010}{053703}.

\bibitem{Kroll:2010}
\Name{Kroll M., Schablinski J., Block D. \and Piel A.} \REVIEW{Physics of
  Plasmas }{17}{2010}{013702}.

\bibitem{Reichhardt:2004}
\Name{Olson~Reichhardt C.~J., Reichhardt C. \and Bishop A.~R.} \REVIEW{Phys.
  Rev. Lett. }{92}{2004}{016801}.

\bibitem{Shlesinger:1993}
\Name{Shlesinger M.~F., Zaslavsky G.~M., Klafter J. \and Morfill G.}
  \REVIEW{Nature }{363}{1993}{31}.

\bibitem{Feldman:1982}
\Name{Feldman L.~C., Mayer J.~W. \and Picraux S.~T.} \Book{Materials analysis
  by ion Channeling} (Academic Press, New York) 1982.

\end{thebibliography}
\end{document}